\documentclass[12pt]{iopart}
\usepackage{iopams}
\usepackage{graphicx}
\usepackage{cite}
\begin{document}

\title[]{Geometric Influence of High Curvature on the Performance of Thin Film Transistors}

\author{Prasenjit Bhattacharya$^{+}$, Aswathi Nair$^{+}$}
\address{Department of Instrumentation and Applied Physics, Indian Institute of Science, India}
\address{$^{+}$ These authors contributed equally to this work}

\author{Sanjiv Sambandan}
\address{Department of Instrumentation and Applied Physics, Indian Institute of Science, India}
\address{Department of Engineering, University of Cambridge, UK}

\ead{ss698@cam.ac.uk}
\vspace{10pt}
\begin{indented}
\item[]August 2018
\end{indented}

\begin{abstract}
The development of thin film transistor (TFTs) based integrated circuits on flexible substrates promise interesting approaches to human interface systems. Recently TFTs have been fabricated on textured surfaces such as textiles, paper, artificially corrugated or dimpled substrates, threads and fibers. This can result in the TFTs metal-insulator-semiconductor (MIS) stack being significantly distorted from a planar zero-curvature geometry to having non zero and potentially high curvature. Although the direct deposition on textured surfaces do not result in mechanical stress (as for example in bending, buckling or wrinkling), the geometry of high curvature can significantly influence the current voltage characteristics of the TFT. Here we present a closed form analytical model describing the geometrical impact of high curvature on the electrical performance of the TFT. Models are obtained from the solution to the Poisson-Boltzmann equation in polar co-ordinates and are verified using TCAD simulations. The techniques to generalize the results to adapt to any texture and semiconducting material are also discussed with experimental verification. This work forms an analytical basis to understand the impact of curvature on the electrostatics of the MIS stack and the current voltage characteristics of the TFT.
\end{abstract}

%
%
%
%
%

\section{Introduction}
Thin film transistors (TFTs) on flexible substrates promise applications such as bendable sensors and actuators \cite{flex0, flex1, flex2, flex3, flex4, flex5, flex6, flex7, flex8, flex9, flex10}, flexible energy harvesters \cite{energy1, energy2, energy3, energy4} and wearable devices for health diagnostics \cite{wear0, wear1, wear2, wear3, wear4, wear5}. In the recent past, there has been a drive towards developing TFTs on rough surfaces such as textiles \cite{TFTonthread3, TFTonthread4, TFTonthread5}, threads and fibers \cite{TFTonthread1, TFTonthread2}, paper \cite{TFTonpaper1, TFTonpaper2, TFTonpaper3, TFTonpaper4, TFTonpaper5,  TFTonpaper6} and artificially corrugated surfaces \cite{sambandan2012influence, aljada2012structured, nair2017modulating, amalraj2014influence}. In such cases, the metal-insulator-semiconductor (MIS) stack of the TFT may experience significant geometric curvature. High curvature of the MIS stack would also occur when TFTs on flexible substrates are bent significantly \cite{bending1, gleskova2002electrical, sekitani2005bending, van2013high, munzenrieder2012flexible, kim2012mechanical, liao2016effect, jamshidi2008effects, bending3, bending4, bending6}. However, while the former case i.e. the deposition of TFTs on non planar surfaces, only brings in a geometric effect, the latter case of bending the TFT involves both, mechanical stress as well as a geometric effect. This work is focussed on the effect of high curvature geometry alone and not the effect of mechanical stress.

High curvature geometry of the MIS stack results in changing the gate electric field distribution from nearly uniform to non-uniform. The importance of geometry on the performance of TFTs is seen from experiments performed on TFTs deposited on artificially engineered rough surfaces \cite{sambandan2012influence, nair2017modulating, aljada2012structured, amalraj2014influence}. In these experiments, the gate metal of the TFT was undulated in both convex and concave fashions by the use of corrugations or dimples. Since the semiconductor and insulator were deposited conformably on top of this gate metal, the MIS stack did not experience mechanical stress and only experienced the effect of curvature. Experiments showed significant variations in electrical performance in the TFT thereby defining the role of geometry alone to be significant in itself. Corrugated gate a-Si:H TFTs showed almost 200-300\% increase in transconductance as compared to planar TFT \cite{sambandan2012influence}. Upto $\pm$40\% variation in transconductance was observed depending on the texture orientation as compared to conventional planar gate TFTs in a-Si:H technology. Demonstrations of high gain common-source amplifiers using textured TFTs have also been reported \cite{nair2017modulating}. Patterning the gate with dimples in organic (P3HT) TFTs showed almost 400\% improvement in current \cite{aljada2012structured}. Therefore, the geometrical impact of curvature on the performance of TFTs is important.

The aim of this work is the development of an analytical model to explain the purely geometrical impact of curvature on electrostatics of the MIS stack and the performance of the TFT. This is accomplished by obtaining the closed form solution to the Poisson-Boltzmann's equation in polar coordinates for positive and negative curvature. We demonstrate the occurrence of a pleasant symmetry in the positive and negative curvature case with the TFT chracteristics being strongly determined by the effective gate capacitance. Models are compared to TCAD simulations and then generalized and compared with experimental data for a corrugated gate TFTs.

\begin{figure}
\centering
\includegraphics[width=4in]{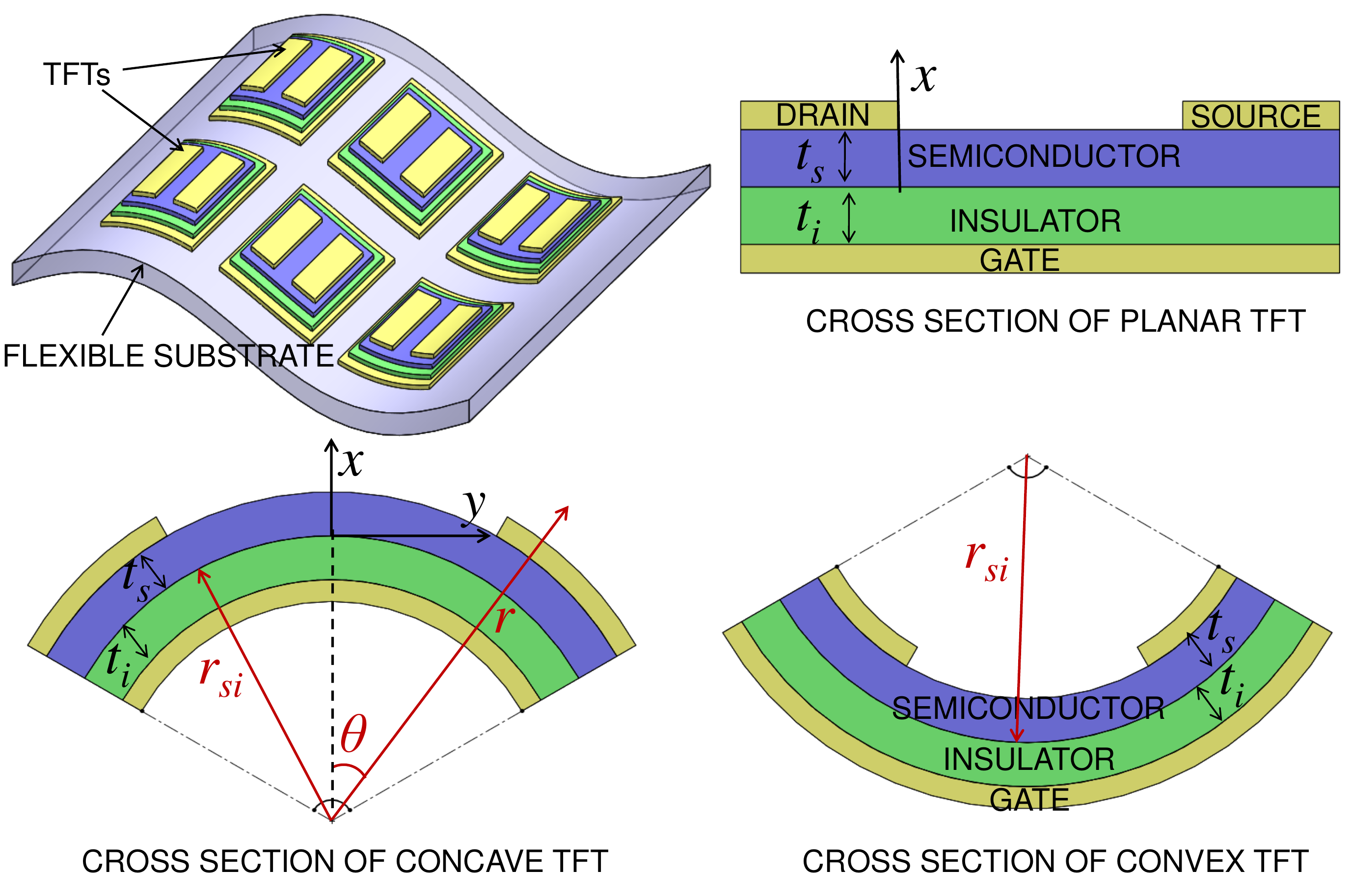}
\caption{Geometrical definitions and \textbf{coordinate} systems (Cartesian $(x,y)$ and polar $(r,\theta)$) used. Cross-section of a TFT with no curvature (planar MIS stack), concave curvature and convex curvature.}
\label{fig1}
\end{figure}

\section{Definitions and Assumptions}
\subsection{Device Physics}
The disordered nature of the semiconductor in a TFT results in the appearance of localized states in the energy bandgap. These can be classified as deep states that are present around midgap and tail states that are present close to the mobility edge. In most inorganic disordered semiconductors (eg. amorphous hydrogenated silicon (a-Si:H) \cite{shur1984physics}, amorphous metal oxides \cite{chen2009density}), the states in the gap are observed to have an exponential distribution with energy.  

For this analysis, we consider n-type TFTs with the electrostatics defined by the Poisson equation as $\nabla^2\varphi=q(n_{t}+n_{f})/\epsilon_{s}$ where $\varphi$ is the potential profile in the semiconductor measured with reference to the back channel (bulk) potential, $q$ the magnitude of the electron charge, $\epsilon_{s}$ the semiconductor permittivity and $n_{t}$, $n_{f}$ are the trapped and free carrier concentration, respectively. To simplify the analysis and preserve physical intuition we represent the contribution of traps at all energies and the effect of the free carrier concentration by defining an effective carrier concentration, $n_{e}$, and an effective characteristic voltage, $\phi_{e}$, and write the Poisson equation as
\begin{equation}
\label{electrostatics}
\nabla^2\varphi=q n_{e}/\epsilon_{s}=q n_{e0}e^{\varphi/\phi_{e}}/\epsilon_{s}
\end{equation}
with $n_{e0}$ being the effective carrier concentration at flat band. The parameters $n_{e}$ and $\phi_{e}$ are influenced by the trapped and free carriers and tend to take values of the specie dominating the physics. This simplifies the analysis without much loss of accuracy and is justified by the simulations and experiments. 

Although the electrostatics of the device is determined by $n_{e}$, the current voltage (IV) characteristics is determined by the free carrier concentration, $n_{f}$. 

\subsection{MIS Geometry}
The geometry of the MIS structure used for analysis is shown in Fig. 1. The MIS has an insulator of thickness $t_{i}$ and semiconductor of thickness $t_{s}$. We define the radius of curvature of the insulator-semiconductor interface to be $r_{si}$. For zero curvature the MIS stack is planar,  $r_{si}=\infty$ and a Cartesian coordinate system $(x,y)$ is used for analysis with the vector $x$ pointing along the direction from the insulator-semiconductor interface ($x=0$) to bulk. When the MIS stack has curvature, a polar coordinate system $(r,\theta)$ is used. The curved MIS stack is said to be concave or convex as shown in Fig. 1. For simulations the arc length of the insulator-semiconductor interface was always kept constant. 

\subsection{Insulator Capacitance Per Unit Area}
We also define an effective insulator capacitance per unit area $C_{i0}$ for the MIS stack with zero curvature, $C_{i+}$ for the case of concave curvature and $C_{i-}$ for the case of convex curvature. It is seen that, $C_{i0}=\epsilon_{i}/t_{i}$, $C_{i+}=\epsilon_{i}/(r_{si}\mbox{ln}(r_{si}/(r_{si}-t_{i})))$ and $C_{i-}=\epsilon_{i}/(r_{si}\mbox{ln}((r_{si}+t_{i})/r_{si}))$, where $\epsilon_{i}$ is the insulator permittivity.

\section{Electrostatics}
\subsection{Zero Curvature (or Planar)}
We first consider the case when the MIS stack has no curvature. This case has been very well investigated \cite{chen2016aphysics}-\cite{deng2014anexplicit}. Poisson's equation is given by (\ref{electrostatics}) and is solved to identify the electrostatics. Using the boundary condition $\xi=0$ at $\varphi=0$ and defining a characteristic length $l_{e}=((2\epsilon_{s}\phi_{e})/(qn_{e0}))^{1/2}$, the electric field ($\xi$) in this case can be shown to vary as 
\begin{equation}
\label{xi_planar}
\xi(\varphi)=\frac{2\phi_{e}}{l_{e}}(e^{\varphi/\phi_{e}}-1)^{1/2}
\end{equation}

Defining the surface potential $\varphi(x=0)=\varphi_{s}$, the second boundary condition can be set as $C_{i0}(V_{gs}-V_{fb}-\varphi_{s})=\epsilon_{s}\xi(\varphi=\varphi_{s})$, with $V_{fb}$ being the flat-band voltage and $V_{gs}$ the gate to source voltage. In strong above threshold operation with $e^{\varphi_{s}/\phi_{e}}>>1$, the solution to the boundary condition is given in terms of the zero order Lambert-W function, $\textbf{W}_{0}$, as
\begin{eqnarray}
\label{phis_planarW}
\varphi_{s}&=&V_{gs}-V_{fb}-2\phi_{e}\textbf{W}_0\left(\frac{\epsilon_{s}/l_{e}}{C_{i0}}e^{(V_{gs}-V_{fb})/2\phi_{e}}\right)\nonumber \\
&\approx& 2\phi_{e}\ln\left(\frac{C_{i0}(V_{gs}-V_{fb})}{qn_{e0}l_{e}}\right)
\end{eqnarray}
Here, the term $\epsilon_{s}/l_{e}$ can be imagined to be a characteristic trap capacitance per unit area. The approximated form of the above expression is obtained by approximating $\textbf{W}_{0}(.)=\mbox{ln}(.)-\mbox{ln}(\mbox{ln}(.))$ for large $V_{gs}$ \cite{ortiz2003exact}. 

The potential profile, $\varphi(x)$, is obtained using (\ref{xi_planar}) along with the boundary condition $\varphi=\varphi_{s}$ at $x=0$,
\begin{equation}
\label{phi_planar_final}
\varphi(x) \approx 2\phi_{e}\ln\left(\mbox{sec}\left(\left \vert \sec^{-1}(e^{\varphi_s/2\phi_{e}})\right \vert -\frac{x}{l_{e}}\right)\right)
\end{equation}

\subsection{Non Zero Curvature}
In the presence of curvature, $r_{si}$ is finite and the Poisson equation can be defined in polar coordinates $(r,\theta)$ as,
\begin{equation}
\label{poisson_polar}
\frac{d^{2}\varphi}{dr^{2}}+\frac{1}{r}\frac{d\varphi}{dr}=\frac{qn_{e0}}{\epsilon_{s}}e^{\varphi/\phi_{e}}
\end{equation}
From the derivation shown in the Supplementary Information, the radial variation of the potential is of the form 
\begin{equation}
\varphi(r)=2\phi_{e}\mbox{ln}\left(\frac{\kappa l_{e}}{r}\mbox{sec}\left(\kappa\mbox{ln}\left(\frac{r}{l_{b}}\right)\right)\right)
\label{phir_polar}
\end{equation}
Here $\kappa$ is a dimensionless coefficient and $l_{b}$ is a characteristic length. The radial component of the electric field can be shown to vary as
\begin{equation}
\label{xi_polar}
\xi=\frac{2\phi_{e}}{l_{e}}\left(\pm\left(e^{\varphi/\phi_{e}}-\left(\frac{\kappa l_{e}}{r}\right)^{2}\right)^{1/2}+\frac{l_{e}}{r}\right)
\end{equation}
with the $+$ for the concave case and $-$ for the convex case. At $r=r_{si}$, $\varphi=\varphi_{s}$ with $\varphi_{s}$ being the surface potential. This sets the boundary condition $C_{i\pm}(V_{gs}-V_{fb}-\varphi_{s})=\pm\epsilon_{s}\xi(\varphi=\varphi_{s})$, with the  $+$ for the concave case ($\varphi$ decreases as $r$ increases) and the $-$ for the convex case ($\varphi$ decreases as $r$ decreases). 

Again considering strong above threshold operation, $e^{\varphi_{s}/\phi_{e}}>>\kappa l_{e}/{r_{si}}$, the solution to (\ref{xi_polar}) can also be presented using the Lambert-W function as was done in the planar case as,
\begin{eqnarray}
\label{phis_polarW}
\varphi_{s}&=&V_{gs}-V_{fb}-2\phi_{e}\textbf{W}_{0}\left(\frac{\epsilon_{s}/l_{e}}{C_{i\pm}}e^{(V_{gs}-V_{fb})/2\phi_{e}}\right) \nonumber \\
&\approx& 2\phi_{e}\ln\left({\frac{C_{i\pm}(V_{gs}-V_{fb})}{qn_{e0}l_{e}}}\right)
\end{eqnarray}
Therefore, for all practical purposes the impact of the geometry of curvature on TFT performance is dictated through the cylindrical capacitive component defined by $C_{i\pm}$. This model for $\varphi_{s}$ is validated by a good match with simulations.

To obtain a good numerical estimate for complete variation of $\varphi(r)$ as well as the surface potential, $\varphi_{s}$, we impose another boundary condition permitting the bands to flatten at the back edge of the MIS capacitor. Therefore, $\xi(r=r_{si}\pm t_{s})=0$ and $\varphi(r=r_{si}\pm t_{s})=0$, with the  $+$ for the concave case and the $-$ for the convex case. This condition helps identify the parameter $\kappa$ to be
\begin{equation}
\label{kappa}
\kappa=(((r_{si}\pm t_{s})/l_{e})^2-1)^{1/2}
\end{equation}
Here the $+$ and $-$ signs are used for the concave and convex cases, respectively. Typically $l_{e}\sim 60$ nm and $t_{s}\sim 100$ nm and despite the high curvature under consideration, $r_{si}$ is expected to easily exceed these values with it taking a minimum value of about $200$ nm for this study. Using the parameter $\kappa$ in (\ref{phir_polar}) and setting $\varphi=\varphi_{s}$ at $r=r_{si}$, we can identify the parameter  $l_{b}$. This permits a closed form expression for $\varphi(r)$ as,
\begin{equation}
\label{phi_polarfinal}
\varphi(r)=2\phi_{e}\mbox{ln}\left(\frac{\kappa l_{e}}{r}\mbox{sec}\left(\pm\left |\mbox{sec}^{-1}\left(\frac{r_{si}}{\kappa l_{e}}e^{\varphi_{s}/2\phi_{e}}\right)\right |-\kappa\mbox{ln}\left(\frac{r}{r_{si}}\right)\right)\right)
\end{equation}
The function $\mbox{sec}^{-1}$ yields both negative and positive values for the same argument. The positive and negative values are appropriate for the concave and convex case, respectively. This is enforced as shown in (\ref{phi_polarfinal}). The models developed for the bending case $(r_{si}<\infty)$, must converge to the corresponding models for the planar case $(r_{si}=\infty)$, as $r_{si}$ is increased. This is verified in Appendix I.

\section{TFT I-V Characteristics}
\subsection{Zero Curvature (or Planar)}
For a TFT with an applied drain to source bias voltage, $V_{ds}$, the model for $\varphi_{s}$ still retains its form (Eq. \ref{phis_planarW}) except for the incorporation of the threshold voltage term, $V_{T0}$, and a channel potential term, $V_{ch}$, instead of $V_{fb}$. Therefore, $\varphi_{s}=2\phi_{e}\ln{(C_{i(0,\pm)}(V_{gs}-V_{T0}-V_{ch})/(qn_{e0}l_{e}))}$. The free electron density per unit area, $N_{f}$ is computed by integrating the free electron concentration per unit volume, $n_{f}$, along the thickness of the semiconductor, so that $qN_{f}=\gamma(C_{i\pm}(V_{gs}-V_{T0}-V_{ch}))^{\alpha-1}$, where, $\alpha=2\phi_{e}/\phi_{th}$ and $\gamma=(n_{f0}(qn_{e0}l_{e})^{2-\alpha}/(n_{e0}(\alpha-1)))$ with $\phi_{th}$ being the thermal voltage. Therefore, the I-V characteristics in above threshold operation when the TFT has no curvature can be shown to be
\begin{equation}
\label{Idseff_planar}
I_{ds}=\gamma\mu_{0} C_{i0}^{\alpha-1}(W/\alpha L)\left((V_{gs}-V_{T0})^{\alpha}-(V_{gs}-V_{T0}-V_{ds})^{\alpha}\right)
\end{equation}
with $\mu_0$ being the band mobility, $W$ the channel width and $L$ the channel length. 

\subsection{Non Zero Curvature}
With curvature, the change in the effective insulator capacitance per unit area from $C_{i0}$ to $C_{i\pm}$ also changes the threshold voltage from $V_{T0}$ to $V_{T\pm}$ as threshold voltage is the ratio of the trapped charge to the insulator capacitance per unit area. Therefore, the I-V model for the case with the TFT experiencing curvature (details in Supplementary Information) is given by, 
\begin{equation}
\label{Idseff}
I_{ds}=\gamma\mu_{0} C_{i\pm}^{\alpha-1}(W/\alpha L)\left((V_{gs}-V_{T\pm})^{\alpha}-(V_{gs}-V_{T\pm}-V_{ds})^{\alpha}\right)
\end{equation}
with the  $+$ for the concave case and the $-$ for the convex case.

Therefore, the difference of the I-V characteristics in the zero and non-zero curvature case is primarily due to the difference in the effective capacitance per unit area. The IV model can be calibrated for TFTs with different materials by suitably changing the parameters $\gamma$, $\alpha$, $\mu_0$ and $V_T$. Free or trapped carriers which will dominate the electrostatics in the above-threshold region is dictated by the values of $\alpha$ and $\gamma$.


\begin{figure*}
\centering
\includegraphics[width=6 in]{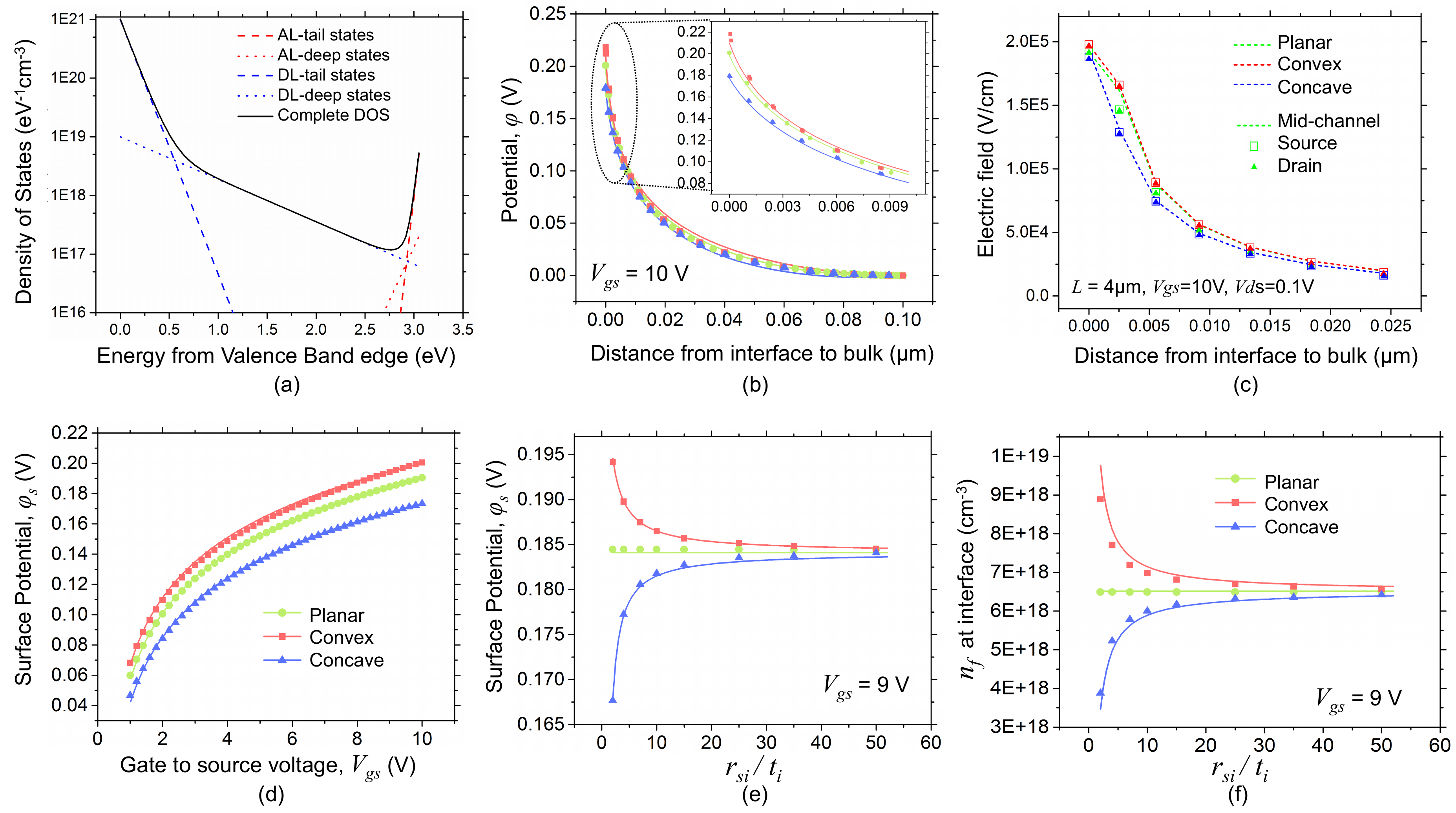}
\caption{Comparison of the electrostatics of the TFT as defined by the analytical model with TCAD simulation. Bold line indicate models and markers indicate simulated data. (a) Density of the states (Properties of a-IGZO) used for simulations (AL-Acceptor like, DL-Donor like). Band-gap=3.05 eV, Effective DoS at conduction-band edge = 5e18 cm$^{-3}$eV$^{-1}$, Density of acceptor-like tail states at conduction-band edge = 5e18 cm$^{-3}$eV$^{-1}$, Characteristic energy of acceptor-like tail states = 30 meV, Density of acceptor-like deep states at conduction-band edge = 2e17 cm$^{-3}$eV$^{-1}$, Characteristic energy of acceptor-like deep states = 120 meV \cite{chen2009density,deng2014anexplicit,bae2013analytical}. (b) Potential variation inside semiconductor from the semiconductor-insulator interface to the bulk using models given in (\ref{phi_planar_final}) and (\ref{phi_polarfinal}), with the inset showing the variation near the interface, (c) Comparison of the electric fields along mid-channel, source edge and drain edge (differentiated by symbols and dashed lines) for different geometries (differentiated by colours) at a channel length $L=4$ $\mu$m. For TFTs with curvature, $r_{si}/t_{si}=10$. All the TFTs were biased at $V_{gs}=10$ V and $V_{ds}=0.1$ V, (d) variation of surface potential with $V_{gs}$ at $r_{si}/t_{i}=2$, (e) variation of surface potential with $r_{si}/t_{i}$ using models (\ref{phis_planarW}) and (\ref{phis_polarW}), (f) variation of $n_{f}$ at interface (given by $n_{f0}e^{\varphi_{s}/\phi_{th}}$) as a function of $r_{si}/t_{i}$.}
\label{fig2}
\end{figure*}

\begin{figure}
\centering
\includegraphics[width=3 in]{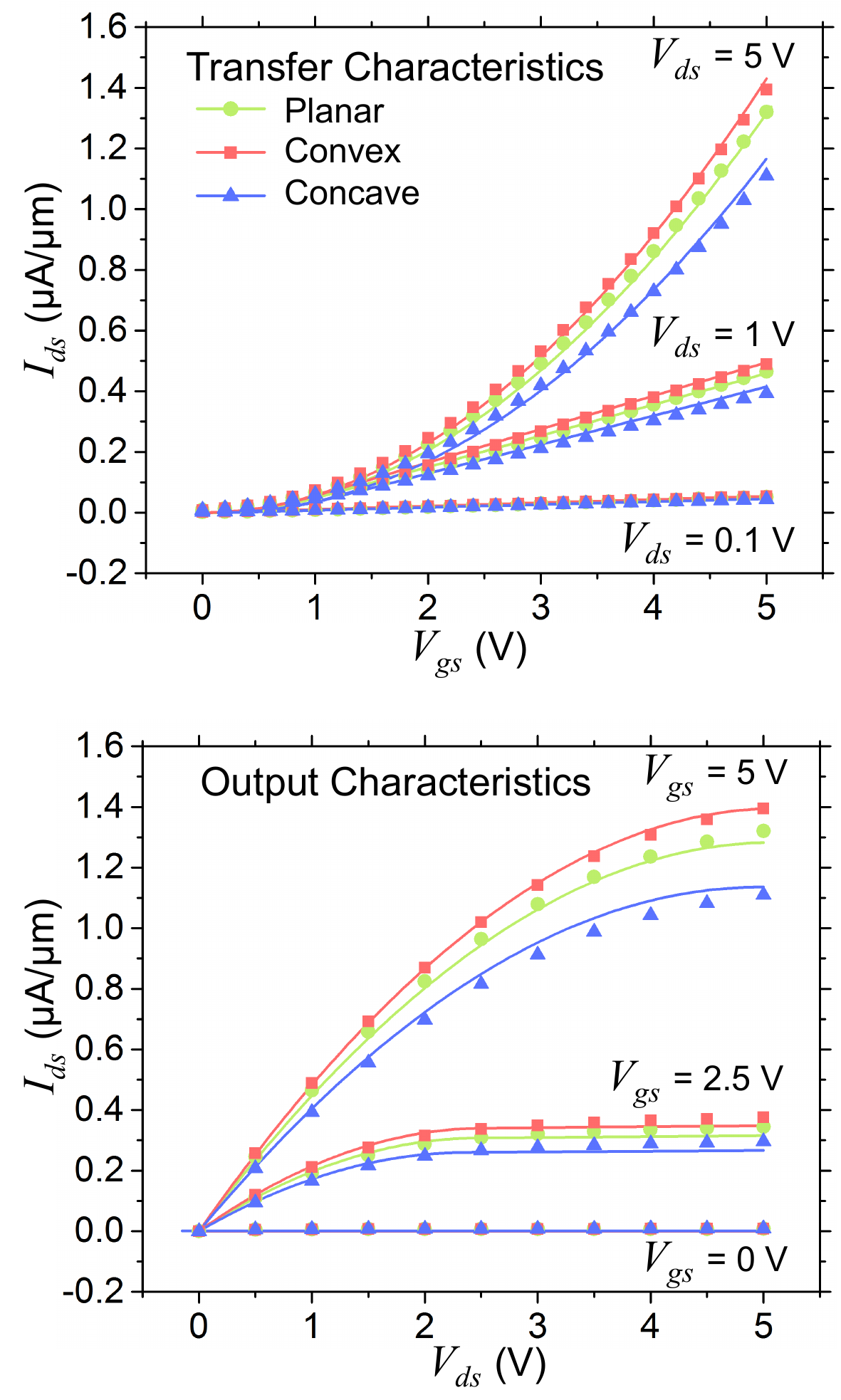}
\caption{Comparison of the current-voltage charactertistics of the TFT (planar, convex and concave IGZO TFTs) as defined by the analytical model ((\ref{Idseff_planar}) and (\ref{Idseff})) with TCAD simulation. Bold line indicate models and markers indicate simulated data.}
\label{fig3}
\end{figure}

\section{TCAD Simulations}
\subsection{Setup}
To validate the models, ATLAS TCAD simulations on cylindrical metal-insulator-semiconductor TFTs were performed. Planar, concave and convex geometries were designed with 100 nm SiO$_2$ insulator, 100 nm a-IGZO semiconductor and Mo gate metal. The default material properties as defined in the simulator were used for the a-IGZO (electron affinity=4.16 eV, electron band mobility = 15 cm$^2$/Vs). The DoS used is as shown in Fig. 2a \cite{chen2009density,deng2014anexplicit,bae2013analytical}.  Simulations for MIS with bending were performed for $r_{si}/t_{i}$ = 2, 4, 7, 10, 15, 25, 35, 50. For MIS simulations the source and drain were grounded and the arc length of the semiconductor-insulator interface was 1.2 $\mu$m for all structures. For TFT simulations, the arc length of the semiconductor-insulator interface was 4.2 $\mu$m, the channel length was 4 $\mu$m (large enough to reduce drain induced barrier lowering) and $r_{si}/t_{i}$ was 7 for all geometries. 

The measurement of $\varphi(r)$ was done using a precisely defined cut-line. A fine mesh was used in the semiconductor while keeping the number of mesh points constant for the different structures for the sake of accuracy. As the influence of the gate voltage at the back channel was negligible, $\varphi_s$ was measured by precisely probing and calculating the difference between the potentials at the interface and the back channel. 

\subsection{Results}
The TCAD simulation results (markers) are compared with analytical models (bold lines) in Fig. 2 (electrostics) and Fig. 3 (current-voltage characteristics). 

The values for $n_{e0}\sim 8\pm 0.1 \times 10^{15}$ cm$^{-3}$ and $\phi_{e}\sim 25.8 \pm 0.7$ mV were estimated by best fit between the model of (\ref{phis_planarW}) and simulation for the case of zero curvature. In the case of IGZO TFTs, calculation of $n_{f}$ and $n_{t}$ using methods described in \cite{leroux1986static} shows that $n_{f}$ is slightly greater than $n_{t}$ in the above-threshold operation. Therefore $\phi_{e}$ is expected to be close to the thermal voltage as corroborated by a good match between analysis and simulation. These parameters were also used for the case of non-zero curvature to obtain good corroboration between model and simulation.

Using the models described in (\ref{phis_planarW}) and (\ref{phis_polarW}), the surface potential of planar, concave and convex TFTs were compared against simulations. The spatial variation of the potential inside the semiconductor for the case of non zero curvature was calculated using (\ref{phi_polarfinal}). Fig. \ref{fig2}(b) and Fig. \ref{fig2}(c) show the spatial variation of $\varphi$ and the electric field (at $V_{gs}=10$V) while moving from the semiconductor-insulator interface to the bulk. The electric field was probed along three cutlines directed radially from the interface towards the bulk at the - mid-channel, edge of source terminal and the edge of drain terminal. Simulation results reveal that the influence of curvature on the electric-field along the three cutlines follow a similar trend for TFTs having $L>4$ $\mu$m (both markers and dashed lines indicate simulations). This justifies the modeling of the electrostatics of the TFT using the polar Poisson-Boltzmann equation as described and without requiring any consideration with regards to the location along the channel length as the source drain edge does not have any different influence with regards to the general trend.

Fig. \ref{fig2}(d) and Fig. \ref{fig2}(e) show the plots of $\varphi_s$ vs $V_{gs}$ (for $r_{si}/t_{i}=2$) and $\varphi_s$ vs $r_{si}/t_{i}$ (at $V_{gs}=9$V), respectively. As predicted by the model, the convex TFTs are observed to have a higher $\varphi_{s}$ as compared to planar devices while the concave TFTs have a lower $\varphi_{s}$ as compared to planar TFTs for the same applied gate bias. As discussed, this is attributed to the variation in effective insulator capacitance per unit area. As $r_{si}$ increases, $\varphi_s$ of the TFTs with curvature approach the case of TFTs with zero curvature. Therefore, although geometry has an impact for any curvature, the influence becomes significant (\textgreater10\%) only when the $r_{si}/t_{i}<20$. This degree of curvature would appear when the substrate or MIS stack buckles, wrinkles or if the TFTs are deposited on fibers.

Fig. \ref{fig2}(f) shows $n_{f}(\varphi=\varphi_{s})$ as a function of $r_{si}/t_{i}$ at $V_{gs}=9$V. Here, $n_{f0}\approx 6.2 \times 10^{15}$ cm$^{-3}$. The model and simulations once again show a reasonably good fit. At a given $V_{gs}$ free electron concentration is higher for the convex curvature case as compared to the zero curvature case. On the other hand, the free carrier concentration for the concave curvature case is lower as compared to the zero curvature case. As expected, this follows a trend similar to $\varphi_{s}$ and is strongly dictated by the variation in gate capacitance. 

Fig. \ref{fig3} compares the transfer and output characteristics for TFTs with zero curvature, convex curvature and concave curvature cases ($r_{si}/t_{i}$=7). At low channel lengths ($\leq 2$ $\mu$m), drain induced barrier lowering was observed. Hence the simulations were performed using a channel length of 4 $\mu$m to study the impact of curvature. The threshold voltage values used to fit the simulated data are $V_{T0}=0.04$V, $V_{T-}=0$V, $V_{T+}=0.15$V. The models based on \ref{Idseff} are compared with simulations and show a reasonably good fit. The best fit was achieved for $\gamma=1$ and $\alpha=2$, which indicates the domination of free carriers in the above threshold regime.

\begin{figure*}[!t]
\centering
\includegraphics[width=6in]{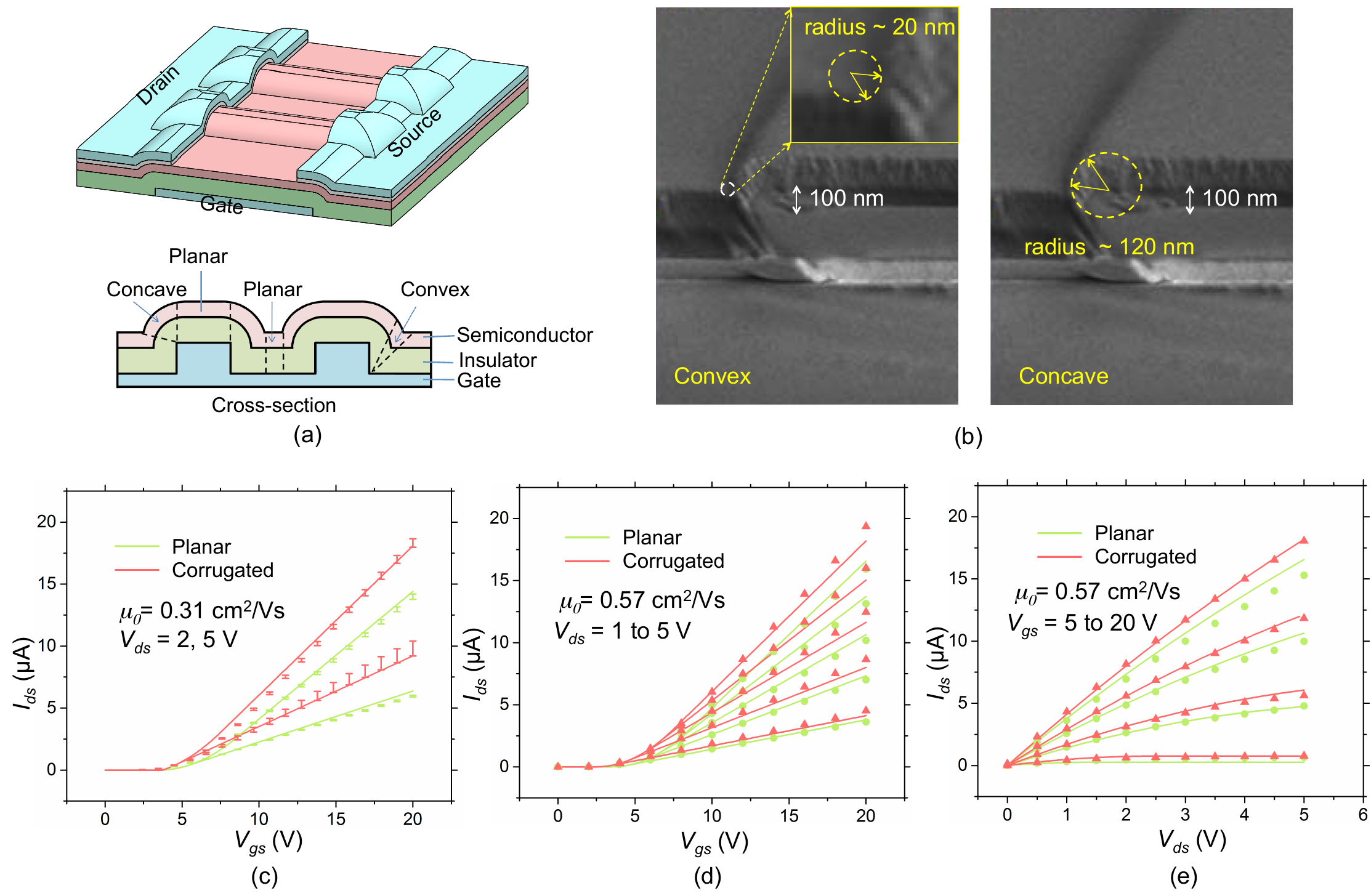}
\caption{(a) Structure of the corrugated gate TFT. (b) SEM image of the cross-section of the corrugated TFT (c) Transfer characteristics along with error bars for planar and corrugated TFTs at $V_{ds}=2,5$V. Bold lines indicate models and error bars indicate experimental data from \cite{sambandan2012influence}. (d) Transfer characteristics at $V_{ds}=$ 1 to 5 V and (e) Output characteristics at $V_{gs}=$ 5 to 20 V for planar and corrugated gate TFTs. Bold lines indicate models and markers indicate experimental data (circles indicate planar and triangles indicate corrugated TFTs) from \cite{sambandan2012influence}.}
\label{fig4}
\end{figure*}

\section{Application to Experiment and Generalizations}
In this section we detail how the analytical models described above can be used against experiments performed with TFTs of different materials and different surface curvatures. As a test example, we consider previously reported experimental data for an artificially corrugated gate a-Si:H TFT, whose schematic is given in Fig. \ref{fig4}(a) \cite{sambandan2012influence}.

\subsection{Generalizations Regarding Geometry}
As seen in Fig. \ref{fig4}(a), the architecture of the MIS stack is more complicated than a simple curved surface. Such a corrugated architecture can be expected when developing TFTs on fabrics without full planarization. To apply the models described in this work, we would first need to break down the geometry into surfaces of convex, concave and zero curvature. The cross-section of the corrugated TFT of Fig. \ref{fig4}(a) can be considered to be composed of exactly such a combination of planar, concave and convex TFTs, with spatially varying radius of curvature. If we define $y$-direction to be along the channel length the $z$-direction be along the channel-width, the drain current per unit width is given by $qN_{f}(y,z)\mu_{0}(dV_{ch}/dy)$. Substituting for $N_{f}$ from the models developed earlier, $I_{ds}/dz=\gamma \mu_{0} [(C_{i(0,\pm)}(V_{gs}-V_{T0}-V_{ch}))]^{\alpha-1}(dV_{ch}/dy)$. The term $\gamma \mu_{0}$ can be considered to be the effective mobility. The term $C_{i(0,\pm)}$ takes on its appropriate values (with subscript 0, + or -) in the corresponding regions (of zero, concave or convex curvature, respectively). For simplicity, the threshold voltage is kept constant. Since the corrugations stretch from source to drain and are parallel to each other and to the direction of the channel lengt, the capacitance varies spatially as we traverse the direction of the channel width and remains constant as we traverse the channel length. The TFT is thus considered to be a parallel combination of TFTs with different gate capacitance and of width $dz$. Integrating the above expression along the length from $y=0$ to $y=L$ and then along the channel width from $z=0$ to $z=W$, $I_{ds}=((\gamma \mu_{0})/(\alpha L))(\int_{0}^{W}{C_{i(0,\pm)}^{\alpha-1}}dz)[(V_{gs}-V_{T0})^{\alpha}-(V_{gs}-V_{T0}-V_{ds})^{\alpha}]$. If the capacitance were to be a continuous function it could be directly integrated in the above expression. However for the case of Fig. \ref{fig4}(a), the geometry and the corresponding value of the capacitance needs to be handled piece wise based on the three periodically repeating regions – the concave, the convex and the planar. The geometrical parameters influencing capacitance, insulator thickness, height of the side wall, corrugation pitch, radius of convex (20 nm) and concave (120 nm) regions, were extracted (likely or approximate values) from the SEM image of the cross section shown in Fig. \ref{fig4}(b). The integral $\int_{0}^{W}{C_{i(0,\pm)}^{\alpha-1}}dz$ was estimated by the piece-wise summation of the separate capacitances.

Although not used in the case of Fig. \ref{fig4}, the current voltage characteristics when the corrugations are parallel to the channel width can be shown to be $I_{ds}=(\gamma \mu_{0}W/\alpha)(\int_{0}^{L}{C_{i(0,\pm)}^{1-\alpha}}dy)^{-1}[(V_{gs}-V_{T0})^{\alpha}-(V_{gs}-V_{T0}-V_{ds})^{\alpha}]$. 

\subsection{Generalizations Regarding Materials}
Before computing the current, it is also important to consider the material used for the TFT. While the simulations were performed using a-GIZO TFTs, the experiments of Fig. \ref{fig4} use a-Si:H as the semiconductor. The material properties would be reflected in the values of parameters $\alpha=2\phi_{e}/\phi_{th}$ and $\gamma=n_{f0}(qn_{e0}l_{e})^{2-\alpha}/(n_{e0}(\alpha-1))$ with $l_{e}=(2\epsilon_{s}\phi_{e}/qn_{e0})^{1/2}$. 

The value of $\phi_{e}$ depends on the region of operation and takes into account the cumulative effect of free carriers and trapped carriers (to aid a single exponential based Poisson equation). The dominant of the two will define $\phi_{e}$. If the free carriers dominate, $\phi_{e}=\phi_{th}\approx 25.8$ mV at 300 K. In a-Si:H, the acceptor like tail states have a characteristic temperature of 325 K and hence a characteristic voltage of about 28 mV \cite{shur1984physics}. Therefore, in above threshold operation in a-Si:H TFTs, $\phi_{e}$ can be expected to take on a value between 25.8 mV and 28 mV. For modeling the experimental data of Fig. \ref{fig4}, we define $\phi_{e} \approx \phi_{th}$. Therefore, $\alpha \approx 2$ and $\gamma \approx n_{f0}/n_{e0}$.

The parameter $n_{e0}$ also takes into account the cumulative effect of free carriers, $n_{f0}$, and trapped carriers. In a-GIZO, the free carriers dominated the trap carriers and $n_{f0}=n_{e0}$ and $\gamma=1$. In a-Si:H \cite{shur1984physics}, the carriers in tail states are quite significant and define $n_{e0}$. In a-Si:H, with the band gap of 1.8 eV, effective density of conduction band states of 2.5e20 /cc, exponential acceptor like tail states with density at conduction band mobility edge and characteristic energy of 1e22 /cc and 0.028 eV respectively, acceptor like deep states with density of 1.5e15 /cc at energy 0.62 eV below conduction band mobility edge and with flat band defined at the point when the fermi level is 0.76 eV below the mobility edge, $\gamma=n_{f0}/n_{e0}\approx 0.02$.

The parameter $\gamma$ can be used to compute the effective mobility. With a band mobility in a-Si:H being $\mu_{0}$=20 cm$^{2}$/Vs, the effective mobility is $\gamma \mu_{0}\approx$ 0.4 cm$^{2}$/Vs. 

\subsection{Model Parameters and Comparison with Experiment}
Based on the above discussion, the values for model parameters used to fit the experimental data are $\mu_0=0.44 \pm 0.13$ cm$^2$/Vs, $\epsilon_{i}=7.5$, $\gamma=1$, $\alpha=2$, $V_{T0}=3.5 \pm 1$ V. The model was also corrected for leakage currents as reported in \cite{sambandan2012influence}. Fig. \ref{fig4}(c)-(e) show good fitting between the I-V models (bold line) and the experimental data (markers) given in \cite{sambandan2012influence}. For best fit $\mu_{0}$=0.31 cm$^{2}$/Vs was chosen for Fig. 4c and $\mu_{0}$=0.57 cm$^{2}$/Vs was chosen for Fig. 4d and Fig. 4e.

\section{Conclusion}
This work presented a closed form analytical model describing the geometrical impact of high curvature on the electrostatics of the MIS stack and the performance of the TFT. The models for TFTs experiencing curvature were obtained from the solution to the Poisson-Boltzmann equation in polar co-ordinates. These models were compared with the case of planar TFTs having no curvature and verified using detailed TCAD simulations. The methods to generalize the results was also discussed using the example of corrugated a-Si:H TFTs. This work forms an analytical basis to understand the behavior of substrate geometry on TFTs, for example with regards to TFTs deposited on rough or textured surfaces such as textiles or paper, surfaces with high curvatures such as fibers or wires and on substrates experiencing buckling, wrinkling or significant bending.

\section*{Acknowledgement}
The authors acknowledge the Department of Science and Technology for funding under the grant DST IMPRINT Program, Grant No. 7969. Sanjiv Sambandan also acknowledges the DBT-Cambridge Lectureship program for permitting a joint faculty appointment at the Indian Institute of Science, India and the University of Cambridge, UK.

\section*{Appendix I}
We perform a check on all models developed by subjecting them to a variation in $r_{si}$, i.e.,  the models for $r_{si}<\infty$ (non-zero curvature) must converge with the models for $r_{si}=\infty$ (zero curvature) as $r_{si}$ is increased. Firstly as $r_{si}$ increases, $t_{i}/r_{si}<<1$ and a Taylor expansion of the logarithm shows that $C_{i+}$ and $C_{i-}$ converge to $C_{i0}=\epsilon_{i}/t_{i}$. Therefore, models for the I-V characteristics and $\varphi_{s}$ for the non-zero curvature case approach the models for the zero curvature case as $r_{si}\rightarrow \infty$. Secondly we consider the parameter $\kappa$ (\ref{kappa}). If $r_{si}>>(t_{s},l_{e})$, $\kappa \approx r_{si}/l_{e}$. Using this, the electric field at the semiconductor-insulator interface (i.e. at $r=r_{si}$ and $\varphi=\varphi_{s}$) in the non-zero curvature case approaches the field at the interface in the zero curvature case. Next, considering the spatial variation of $\xi$ and $\varphi$ we note that the polar coordinate system $(r,\theta)$ maps to the Cartesian coordinate $x$ as $x=\pm(r- r_{si})\mbox{cos}(\theta)$ with the interface at $x=0$ and $r=r_{si}$. As all parameters in the non-zero curvature case are invariant with $\theta$, we set $\theta=0$ and use $r=r_{si}\pm x$ in (\ref{xi_polar}) and (\ref{phi_polarfinal}). As $r_{si}$ becomes very large,  firstly $\kappa l_{e} \approx r_{si}$, secondly $l_{e}<<r_{si}$ and finally since $x$ spans from $0$ to $t_{s}$, $x/r_{si}<<1$. Therefore, $r_{si}/(r_{si}\pm x)\approx 1$ and $\kappa\mbox{ln}(1\pm(x/r_{si}))\approx \pm\kappa x/r_{si}=\pm x/l_{e}$. Using $r=r_{si}\pm x$ along with these approximations we find that (\ref{xi_polar}) and (\ref{phi_polarfinal}) converge to (\ref{xi_planar}) and (\ref{phi_planar_final}), respectively.

\section*{References}
\bibliographystyle{iopart-num}
\bibliography{refbending}

\end{document}